An anthropomorphic thyroid phantom for ultrasound-guided radiofrequency ablation of nodules

Running title: A thyroid nodule thermal ablation phantom.

Tim Boers[1*], Wyger Brink[3], Leonardo Bianchi[1,2], Paola Saccomandi[2], Johan van Hespen[1], Germen Wennemars[3] ,Sicco Braak[4], Michel Versluis[5], and Srirang Manohar[1]

[1]Multi-Modality Medical Imaging group, TechMed Centre, University of Twente, Enschede, the Netherlands

[2]Department of Mechanical Engineering, Politecnico di Milano, Milan, Italy

[3]Magnetic Detection and Imaging group, TechMed Centre, University of Twente, Enschede, the Netherlands

[4]Department of Radiology, Ziekenhuisgroep Twente, Almelo, the Netherlands

[5]Physics of Fluids group, TechMed Centre, University of Twente, Enschede, the Netherlands

**\* Corresponding author:**

Tim Boers

t.boers@utwente.nl

Drienerlolaan 5, 7522 NB, Enschede

Room: Technohal 2386

Phone number: +31 53 489 1777

Abstract

Background: Needle-based procedures such as fine needle aspiration (FNA) and thermal ablation, are often applied for thyroid nodule diagnosis and therapeutic purposes, respectively. With blood vessels and nerves nearby, these procedures can pose risks in damaging surrounding critical structures.

Purpose: The development and validation of innovative strategies to manage these risks require a test object with well-characterized physical properties. For this work, we focus on the application of ultrasound-guided thermal radio-frequency ablation (RFA).

Methods: We have developed an anthropomorphic phantom mimicking the thyroid and surrounding anatomical and physiological structures that are relevant to ultrasound-guided thermal ablation. The phantom was composed of a mixture of polyacrylamide, water, and egg white extract and was cast using molds in multiple steps. The thermal, acoustical, and electrical characteristics were experimentally validated. The ablation zones were analyzed via non-destructive $T_2$-weighted MRI scans utilizing the relaxometry changes of coagulated egg albumen, and the temperature distribution was monitored using an array of fiber Bragg sensors.

Results: The physical properties of the phantom were verified both on ultrasound as well as its response to thermal ablation. The final temperature achieved (92°C), the median percentage of the nodule ablated (82.1%), the median volume ablated outside the nodule (0.8 mL), and the median number of critical structures affected (0) were quantified.

Conclusion: An anthropomorphic phantom that can provide a realistic model for development and training in ultrasound-guided needle-based thermal interventions for thyroid nodules has been presented. In the future, this model can also be extended to novel needle-based diagnostic procedures.

Keywords:        Thyroid phantom, ultrasound, MRI, radiofrequency ablation, polyacrylamide gel

1.   Introduction

Thyroid nodules are tumors in the thyroid parenchyma, which are often radiologically distinguishable[1]. Nodules can be found in up to 67% of the adult population[2,3], and the prevalence appears to be increasing, from 21.53% to 29.29% in the last decade[4]. In 90-95% of the cases, the nodules are found to be benign[5]. In thyroid nodule management, ultrasound-guided minimally invasive approaches are becoming increasingly prevalent, the majority of them being biopsies and, specifically for benign nodules, thermal ablations[6]. While biopsies and ablations offer valuable strategies for diagnosis and therapy respectively, the efficacy of these interventions can be further improved. For instance, up to 24.1% of patients who undergo thermal ablations such as radiofrequency ablation (RFA) can suffer from regrowth of the nodule due to incomplete ablation[7,8]. These issues may originate from incorrect placement of the RF electrode due to limited imaging quality,  due to limited experience of, and choice of technique by the operator[8,9]. Also, limited access of the RF electrode due to surrounding critical structures requires a safety margin[7,8].

There is a need and an ongoing effort[10,11] to improve current approaches in needle-based diagnosis and interventions on thyroid nodules. Testing and iterative optimization of the technique is not possible directly on patients, due to its burden and due to ethical considerations. A common approach is to perform initial tests and optimization in a laboratory setting on inanimate objects called phantoms[12]. These phantoms can vary from being relatively simple[13] covering basic anatomy to fairly complex allowing for tuning the material characteristics[14] and mimicking more realistic anatomical geometries[14] as well as incorporating physiologic parameters such as blood flow[15,16]. Several thyroid phantoms, intended for training and research into biopsies and thermal ablations, are described in the literature. The main goal for the biopsy phantoms was training[17,18]. Of these phantoms, one was produced in-house and was fully anthropomorphic[17], another commercial phantom offered a semi-anthropomorphic visualization[18]. For the thermal ablation studies, one phantom was produced for an MRI-compatible high-intensity focused ultrasound system[19]. Another phantom was produced for an RFA study comparing various ultrasound transducers[13].

The materials used in phantoms vary as well, where the literature is reporting water-based, oil-based as well as plastic materials[20]. Specifically for phantoms intended for thermal therapies, thermochromic additives or albumins can be used[14, 21–23], to provide an indication of the maximum temperature achieved in the treatment area. The former approach requires manual slicing and visual or optical analysis of the colored regions as indicative of the ablation zones[13], which provides only cross-sectional information that, combined, merely approximates the ablation zone volume. In the latter approach, the coagulation of albumin produces an irreversible change in $T_2$-relaxation time which can be characterized using Magnetic Resonance Imaging (MRI), facilitating a non-destructive analysis of the ablation zone. A realistic thyroid nodule phantom that can capture the relevant anatomical and physiological aspects and facilitate a non-destructive means of validating these new technologies will expedite the development process.

In this work, we present an anthropomorphic neck phantom suitable for technological development and training in ultrasound-guided RFA-based interventions for thyroid nodules. The phantom combines multiple properties, described in literature separately, into one model. It is based on a polyacrylamide gel with additives to mimic tissue properties for ultrasound propagation, as well as the temperature-sensitive albumin that exhibits changes in $T_2$ MR relaxation time upon coagulation, thereby facilitating non-destructive volumetric analysis of the ablation zone. To mimic the presence of the major blood vessels and their associated heat-sink effect, thin-walled vessels were incorporated in the phantom with water flow at 37°C. Further, the phantom contains an anatomically realistic thyroid with two nodules incorporated in two positions (lateral and caudal in the thyroid lobes), which can be imaged both on ultrasound as well as MRI. The physical properties of the phantom materials were first determined experimentally. The thyroid nodules then underwent RF ablation, guided by ultrasound imaging. The temperature was continuously monitored using fiber Bragg grating sensor technology, and evaluated afterward using a quantitative $T_2$-mapping MRI technique.

## 2. Materials and methods

### 2.1. Phantom morphology, materials, and fabrication protocol

A three-dimensional anthropomorphic thyroid phantom was designed based on a segmentation of *in-vivo* MRI data, obtained from the AAPM RT-MAC Grand Challenge 2019[24,25]. It was designed to contain the trachea, thyroid, thyroid nodules, carotid arteries, internal jugular veins, recurrent laryngeal nerves, and vagal nerves. The dimensions of the internal morphology were based on the average dimensions of the human neck[26–28]. Two thyroid nodules (both near 7.5 mL in volume) were digitally sculpted and merged with the thyroid model.

Three thyroid models were created to represent different nodule topologies: a left lateral and right caudal nodule position, a left caudal and right lateral nodule position, and, finally, both caudal nodule positions. Three-dimensional renderings of the thyroid variations are shown in Fig. 1. The molds, also including the container and lid, were produced using acrylonitrile butadiene styrene (ABS)-filament on a 3D-printer (S5, Ultimaker, Utrecht, the Netherlands) using a 0.4 mm nozzle, 0.1 mm layer thickness, build plate adhesion and supports for 60° or more of overhang.

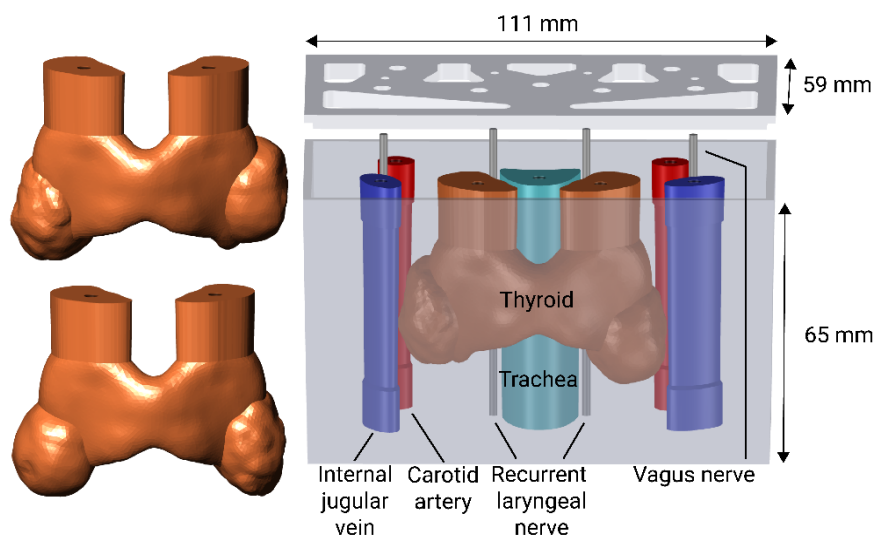

Figure 1. The phantom container holding one of the thyroid molds and surrounding critical structures. Three different nodule topologies were implemented. From a patient perspective the nodule positions are for the left phantom: left lateral and right caudal, for the middle phantom: left caudal and right lateral, and for the right phantom: both caudal.

The anthropomorphic phantom was made of polyacrylamide gel (PAG) with various additives to make it suitable for ultrasound-guided RF ablation studies. Albumin was incorporated as a temperature-sensitive MRI marker, in the form of industrially separated egg white which holds roughly 10.5% albumin[29]. The egg white has a coagulation temperature between 62 and 81°C where 72°C is the main coagulation temperature[30]. The albumin coagulation temperatures are similar to tissue[23]. Silica beads were added to increase the ultrasound scattering coefficient of the gel. A higher concentration was used within the thyroid compartment to produce anatomical contrast within the phantom. Finally, sodium chloride was added to increase the electrical conductivity to physiological levels. The entire ingredient list is shown in Table 1.

Table 1. Ingredients and their percentage in the phantom.

|  | Thyroid % | Body, nodule % |  |
|---|---|---|---|
| Egg white | 50.00 | 50.00 | (v/v) mL |
| (Albumin) | 10.5 | 10.5 | (w/v) g |
| (Water) | 88.50 | 88.50 | (v/v) mL |
| (Sodium chloride, NaCl) | 0.40 | 0.40 | (w/v) g |
| Degassed deionized water | 31.02 | 31.37 | (v/v) mL |
| 40%, 19:1, acrylamide/bis-acrylamide | 17.50 | 17.50 | (v/v) mL |
| Sodium chloride (NaCl) | 0.70 | 0.70 | (w/v) g |
| Silica beads (SiO$_2$) | 0.50 | 0.15 | (w/v) g |
| Ammonium persulfate (APS) | 0.14 | 0.14 | (w/v) g |
| Tetramethylethylenediamine (TEMED) | 0.14 | 0.14 | (v/v) mL |
| **Total volume (%)** | **100.00** | **100.00** |  |

To produce the main phantom body, we first mixed the egg white, degassed deionized water, acrylamide/bis-acrylamide mix (A9926), sodium chloride (S9888), silicon dioxide beads (S5631) and ammonium persulfate (248614), in corresponding order. Tetramethylethylenediamine (T22500) was added shortly before pouring the mixture into the mold, to accelerate the cross-linking process and prevent separation and sedimentation. All ingredients were obtained via a research supplier (Sigma Aldrich, St. Louis, Missouri, USA) with the exception of the deionized water and the industrially separated egg whites (Weko Eiproducten B.V., Ochten, the Netherlands) which were obtained from a local supermarket. The mixture was then poured into a rectangular container 111×59×65mm in size, with the thyroid mold supported from a top plate, to form the main body of the phantom. Negative molds for the nodules were filled using the same mixture. To prevent air bubbles, the phantom bodies were positioned in a vacuum chamber for one minute with gentle wobbling of the chamber to expedite bubble removal. Cross-linking was completed after five minutes, after which the nerve, trachea and vessel molds were gently removed. A small incision was made in the gel above the isthmus, to facilitate the thyroid mold removal without tearing the gel. The nodules were then inserted into the thyroid cavities in their respective nodule positions. Finally, the thyroid mixture (listed in Table 1) was added to fill the remaining cavity and encapsulate the nodules. This was followed by one minute of degassing in the vacuum chamber and five more minutes for the cross-linking to complete. The completed phantom was removed from the container and vacuum wrapped and stored in a refrigerator at 4°C until thermal ablation was performed. The intermediate and final steps of the phantom production process are shown in Fig. 2.

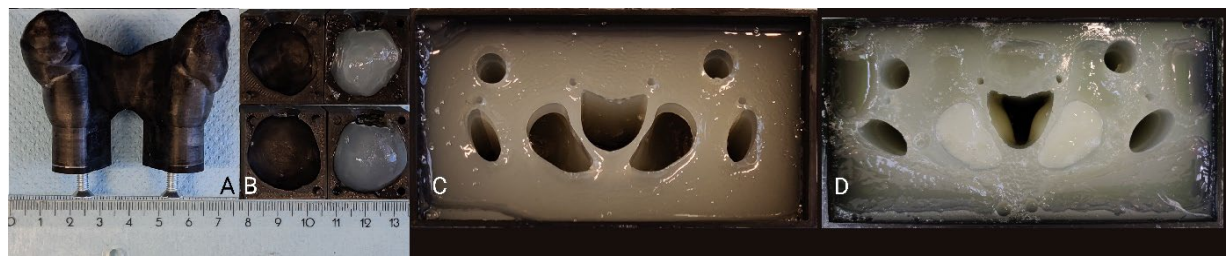

Figure 2. Thyroid model (A) with the cross-linked gel nodules and their molds (B). Top view of the gel block without nodules and thyroid (C). Top view of the finalized phantom in the container, with the nodules and thyroid in place (D).

## 2.2. Phantom characterization

The following physical properties of the phantom were characterized: acoustic attenuation, speed of sound, electrical conductivity, thermal diffusivity, thermal conductivity, specific heat capacity, density, $T_2$ MR relaxation time and its change with temperature.

### 2.2.1. Acoustic attenuation and speed of sound

To determine the acoustic attenuation of the phantom material we used a setup created by Dantuma *et al.*[31]. The setup involved measuring the acoustic transmission across two PAG slabs with different thicknesses using a hydrophone needle and ultrasound receiver, thereby observing the time of arrival and amplitude of the signals. This allows for the determination of the acoustic attenuation coefficient and speed of sound.

### 2.2.2. Electrical conductivity

Two PAG blocks, at room temperature, were ablated eight times each, every time for five seconds, and the electrical impedance after five seconds was reported. The VIVA thyroid RF-ablation device (STARmed Co., Ltd., Goyang, South Korea) was used at 480 kHz. The impedance outcomes were summed and averaged to obtain the electrical conductivity $\sigma$ [S/m]:

$$\sigma = \frac{d}{RA}, \qquad (1)$$

where $d$ is the slab thickness, $R$ is the electrical resistance and the $A$ is the sample cross-section, i.e., the area of the sample that is in contact with the grounding pad.

### 2.2.3. Thermal diffusivity, conductivity and specific heat capacity

The thermal properties are determined as follows. The thermal conductivity $k$ [W/(m·K)] is given by Cooper and Trezek [32]:

$$k = 0.0502 + 0.00577 \times w, \qquad (2)$$

with $w$ the water content, based on the phantom recipe (specified in Table 1).

The specific heat capacity $c_p$ [J/(kg·K)], is given by Cooper and Trezek [32], and Riedel [33,34] as:

$$c_p = 1670 + 25.1 \times w. \qquad (3)$$

Both of the aforementioned equations are valid approximations when the water content is higher than that of fat, which is on average 20% [35–37].

The thermal diffusivity $\alpha$ [m²/s] is given by[38]:

$$\alpha = \frac{k}{\rho c_p}, \qquad (4)$$

where $k$ is the thermal conductivity, $\rho$ is the density and $c_p$ is the specific heat capacity. The volumetric heat capacity $s$ [J/(K·m³)] was determined from:

$$s = \rho c_p. \qquad (5)$$

### 2.2.4. Temperature mapping via MRI-based $T_2$ relaxometry

A non-destructive volumetric approach for evaluation of the ablation zone was developed by leveraging the temperature-sensitive MR properties of albumin, that change irreversibly upon heating. After heating, the attained temperature is imprinted into the phantom material and can be quantified indirectly by means of $T_2$ mapping.

First, the relation between $T_2$ relaxation time and temperature was characterized using a set of spherical PAG phantoms that were heated to reference temperatures. To construct the phantom spheres the polyacrylamide mixture was poured into two trays with 33 spherical molds and left for cross-linking, resulting in 66 PAG spheres with a diameter of 2.8 cm. A single sphere was then packed in a plastic bag and submerged in a temperature-controlled water bath, after inserting a thermocouple in its center. The sphere was then heated for 12 minutes, which ensured reaching a uniform temperature[39], after which the core temperature was recorded, and the sphere was removed from the water bath. The procedure was then repeated for a total of 66 temperature settings, ranging from 48 to 85°C.

Then, $T_2$ mapping of the spheres was performed on a Siemens 1.5T scanner (Aera, Siemens Healthineers, Erlangen, Germany) using a series of 3D $T_2$-weighted fast spin echo sequence with the following settings: field of view = $192 \times 192 \times 64$ mm$^3$, 1 mm$^3$ isotropic voxel size, TR = 1000 ms, TE = 80-400 ms in steps of 80 ms, number of refocusing pulses = 64, echo spacing = 10 ms. The voxel-wise data were fitted to a mono-exponential decay model using MATLAB® (R2022b, Mathworks, Natick, MA, USA). The obtained $T_2$ relaxation times were then combined with corresponding thermocouple temperature readings and a polynomial fit was used to model the transition in $T_2$ time with increasing temperature. Finally, the same MR protocol was then used to characterize the thyroid phantoms after ablation, to generate volumetric temperature maps directly from the $T_2$ maps.

2.2.5. Temperature and fiber Bragg grating sensors

To validate the $T_2$-based temperature maps, we recorded the temperature of the phantom using ten fiber Bragg grating (FBG) sensors. The FBG sensors allow for an independent measurement, wherein both the absolute temperature reached as well as the temperature distribution within the ablation zone could be monitored. Further details on FBG-based thermometry are provided in the works of Korganbayev *et al*. and Bianchi *et al*. [40,41].

We used a highly temperature-resistant polyimide-coated optical fiber inscribed with an array of 10 FBG sensors (FiSens GmbH, Braunschweig, Germany). Each sensor had a grating length of 1 mm and the spacing between consecutive gratings (edge-to-edge distance) was 1 mm. The optical fiber was connected to a portable optical interrogation unit (FiSpec FBG X100, FiSens GmbH, Braunschweig, Germany) to illuminate and recover the reflection spectra from the FBG sensors (wavelength range 790 nm - 890 nm, 10 Hz sampling rate). The temperature evolution and distribution along the fiber were then reconstructed from the optical data using MATLAB® (R2022b, Mathworks, Natick, MA, USA).

Two PAG blocks were ablated, four times each, with the FBG sensors either parallel or orthogonal to the RF electrode (Fig. 3). The two orientations allowed for an evaluation of the temperature distribution within the ablation zone both along the RF electrode as well as across the electrode. In both experiments, the average distance between the RF electrode and the FBG sensors was approximately 3.5 mm. Ablation was performed at a power of 25 W for five minutes. A lower power was used than in the phantom studies (Sec. 2.4) to prevent the formation of gas bubbles that may lead to erroneous readings of the temperatures due to the FBG sensor's cross-sensitivity to strain. The starting temperature of the gel blocks was measured using a k-type thermocouple (Voltcraft 300K Thermometer, Conrad Electronic, Hirschau, Germany) with an accuracy of 2°C, at the start of each ablation.

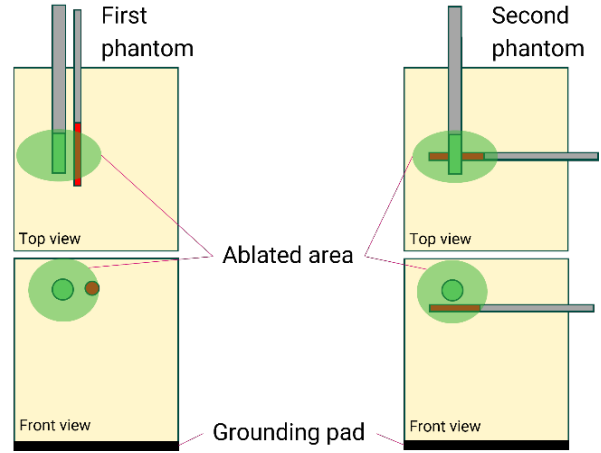

Figure 3. The FBG sensor (red) positioned in parallel with the RF electrode (green) (left) and orthogonal to the electrode (right).

### 2.2.6. Density

For the calculation of the mass density of the gel ($\rho_s$), Archimedes' principle was utilized. Two blocks of PAG were weighed three times in air and three times submerged in water. The temperature of the water was noted during measurements to account for changes in its density. The density is calculated as:

$$\rho_s = \frac{w_a}{w_a - w_w}(\rho_w - \rho_{air}) + \rho_{air} \,, \tag{6}$$

where $w_a, w_w$ are the weights of the gel block in air and water respectively, and $\rho_w, \rho_{air}$ are the densities of water and air, respectively.

### 2.3. Phantom ablation experiments

To mimic the heat-sink effect around veins and arteries, we created a flow setup. This setup consisted of thin-walled pre-stretched balloons (15 cm in length), that were covered in ultrasound gel and inserted in the phantom to serve as arteries and veins (red and blue objects, respectively, in Fig. 4). At the ends of the balloons, 3D printed connectors were inserted. These connectors attach the silicone tubing to a continuous flow pump, which is placed in a temperature-controlled water bath (Anova Precision Pro Cooker, Anova Applied Electronics, Inc., San Francisco, USA), and lead back to the water bath to close the circuit. A schematic drawing of the setup with interconnects is shown in Fig. 4.

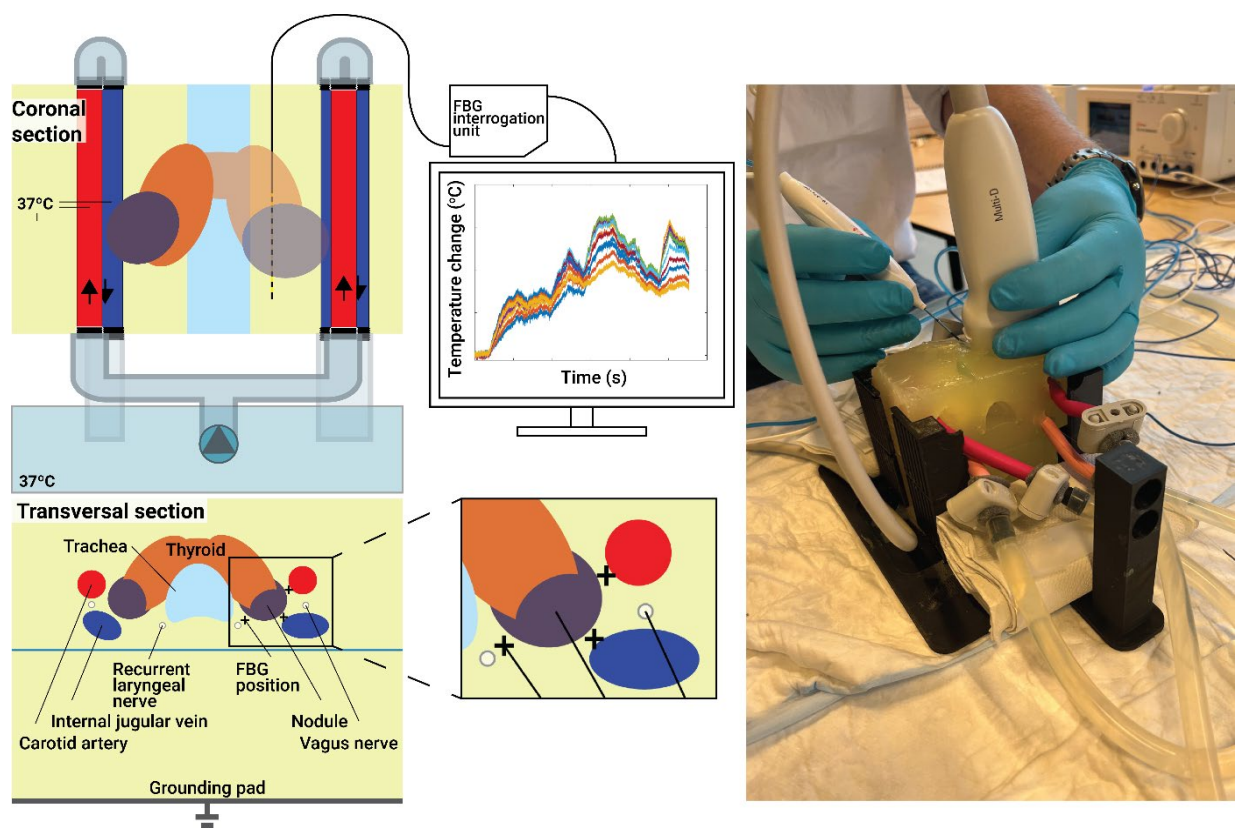

Figure 4. Schematic drawing of the phantom setup. Shown on the left are the phantom body (yellow), silicone tubing and temperature-controlled water bath (transparent blue). Shown in the middle is the FBG (fiber Bragg grating sensors) interrogation unit and readout computer as well as a zoom-in on the FBG positions. Shown on the right is one of the phantoms hooked up to the flow system while being ablated.

In the phantom, one hole per nodule was created with an 18G hollow needle at one of three FBG positions. The three positions are: between the nodule and carotid, between the nodule and internal jugular vein and finally between the nodule and the recurrent laryngeal nerve, as shown in the zoomed-in section in Fig. 4. The optical fiber with FBG sensors was then inserted into this hole.

The temperature of the water bath and phantoms was kept at a temperature of 37°C for one and a half hours before ablation. During ablation, the temperature of the phantom was approximately 22°C on the outside and 37°C on the inside (measured with the k-type thermocouple). Variations throughout the phantom are inevitable due to the required preparation time for the flow setup. The flow was set at 270 mL/min, based on the average flow in the common carotid artery[42]. Three test-phantoms were then ablated according to the moving-shot technique with a transisthmal approach[43] at a power setting of 40 W. The first two phantoms were ablated by a clinician with experience of over 100 ablations performed. The last phantom was ablated by a novice with 0 ablations performed. We used an 18G internally cooled monopolar RF-electrode (7 mm length and 360 degrees active zone) in combination with the VIVA thyroid RF-ablation generator at a pulse repetition frequency of 480 kHz (STARmed Co., Ltd., Goyang, South Korea). The ablation procedure was image-guided using an Acuson S2000 ultrasound system, and a linear transducer (14L5, Siemens Healthineers, Erlangen, Germany) at its center frequency of 11 MHz. After the ablation, 3D $T_2$ maps were obtained and converted into temperature maps to evaluate the extent of the ablation area. Finally, the embedded thyroids were also cut into slices in order to compare the size of the ablation zone by visual inspection with that obtained using MRI.

## 3. Results

### 3.1. Phantom characterization

The results for the phantom material characterization are shown in Table 2, and compared to literature values[37,44]. Overall we find very good agreement with the physical properties reported in the literature, with the exception of the acoustic attenuation and specific heat capacity (and subsequently the volumetric heat capacity) that showed major differences from the literature values.

Table 2. Phantom material characterization results.

|  | Phantom Neck/nodule | Phantom Thyroid | Literature [37,44] | |
|  |  |  | Muscle | Thyroid |
|---|---|---|---|---|
| Acoustic attenuation ($A$) [dB/MHz/cm] | 0.23 | 0.50 | 0.63 | 1.34 |
| Speed of sound ($v$) [m/s] | 1583 | 1586 | 1588 | 1585 |
| Electrical conductivity ($\sigma$) [S/m] | 0.52 | | 0.44 | 0.56 |
| Thermal diffusivity ($\alpha$) [m²/s] | $1.20\text{x}10^{-4}$ | | $1.27\text{x}10^{-4}$ | $1.37\text{x}10^{-4}$ |
| Thermal conductivity ($k$) [W/(m·°C)] | 0.50 | | 0.49 | 0.51 |
| Specific heat capacity ($c_p$) [J/(kg·°C)] | 3826 | | 3544 | 3609 |
| Volumetric heat capacity ($s$) [J/(°C·m³)] | 4132 | | 3863 | 3789 |
| Density ($\rho$) [kg/m³] | 1.08 | 1.08 | 1.09 | 1.05 |

### 3.1.1. Temperature and quantitative T2w MRI validation.

The $T_2$ map obtained for the first 33 phantom spheres is shown in Fig. 5. The sphere borders were excluded from the analysis to avoid contamination from truncation artifacts[45]. The $T_2$ data yielded 66 points which are shown in the graph of Fig. 5. The measurements, repeated over the course of three weeks at one-week intervals, yielded reproducible values with only a small decrease (week zero versus week three) in $T_2$ of 4.4 ms on average, which indicated very stable properties of the (denatured) albumin when considering that the average standard deviation of the $T_2$ was 14.3 ms. This allowed to approximate an average fifth-degree polynomial fit through these points. With this fit, the temperature maps can be derived directly from the $T_2$ maps.

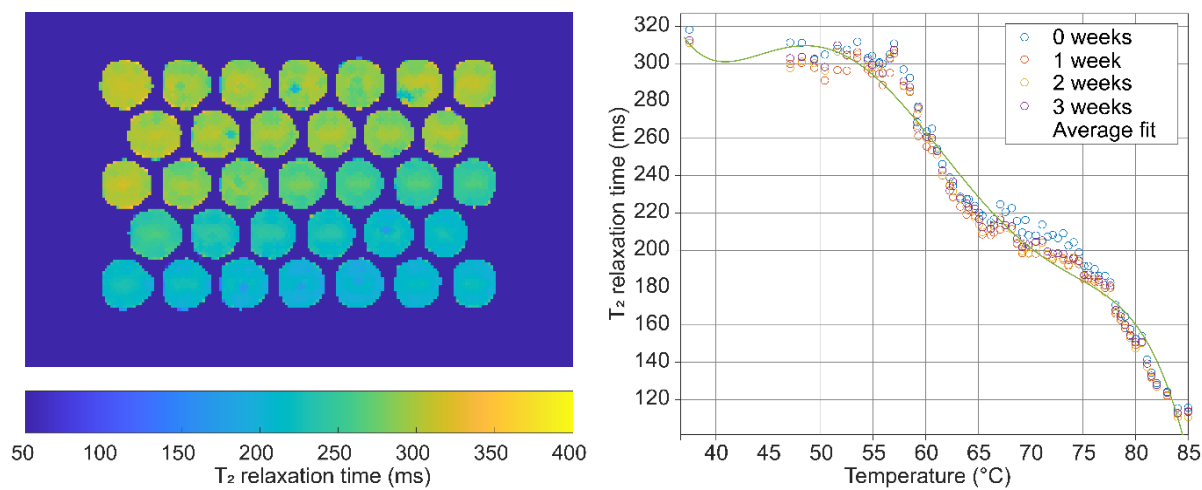

Figure 5. $T_2$ characterization of the phantom material. Shown are a cross-section of the MRI $T_2$ map for 33 out of the 66 phantom spheres (left) and a graph showing the change in $T_2$ relaxation time for the temperature range shown (right). The solid line shows a fifth-degree polynomial fit through the measurements taken at four different time points.

### 3.1.2. MRI temperature and FBG sensors validation

Results from one of the FBG measurements in the two PAG blocks are shown in Fig. 6. The FBG measurement showed no signs of bubble interference, indicated by the monotonic increase and decrease of the temperature. The boxplot in Fig. 6 covers all the results of both the FBG-measured temperatures and the MRI-determined temperatures for the two PAG blocks. Two measurements were excluded due to a too low ablation temperature reached at the measurement site, that did not lead to a sufficiently high degree of coagulation. The mean deviation of the MRI temperature from the FBG measured temperature was 0.8°C.

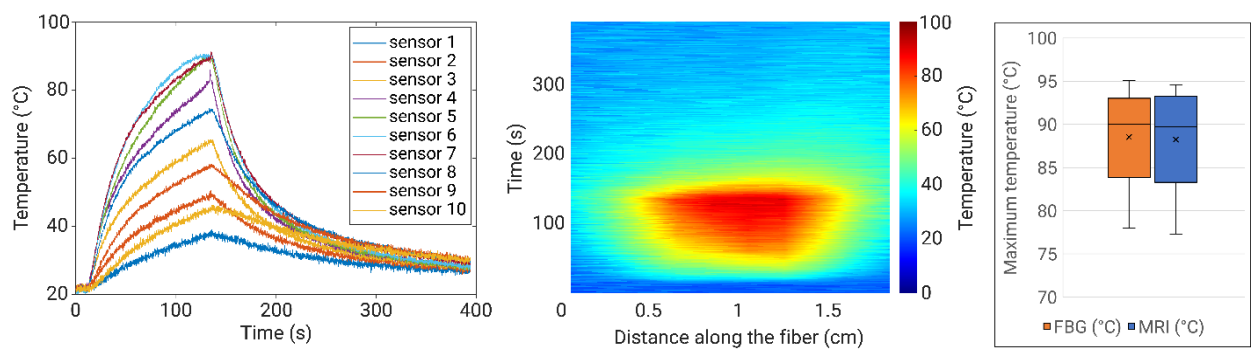

Figure 6. On the left one of the FBG measurements is shown, showing the absolute temperature over time, of the 10 FBG sensors, a maximum temperature of 92.3°C was recorded. In the middle, the temperature recorded using the FBG sensors over time. On the right, a boxplot showing the median and spread of the maximum temperatures recorded for the FBG and MRI measurements.

### 3.2. Phantom ablation experiment

The three phantoms were ablated under ultrasound image guidance as shown in Fig. 4, on the right. Corresponding ultrasound images are shown in Fig. 7, where the relevant structures can be clearly identified, as well as the creation of gas bubbles. A central section of one of the phantoms, corresponding to the $T_2$w MRI cross-section (TE = 240 ms) and the $T_2$ temperature map derived from that cross-section are shown in Fig. 8. These three images show good correspondence of the size of the ablation zone and the limited thickness of the temperature drop-off zone at the edge of the ablation zones. Additionally, a temperature graph for the white dashed line through the nodule is shown in Fig. 8. Please note the consistent temperatures reached throughout the nodule. The FBG sensors readings belonging to the black X position are shown in Fig. 9. The maximum values of those readings are combined with the reading from the MRI temperature map and plotted in the graph on the right in Fig. 9. The graph shows a similar trend for both the FBG and MRI derived temperatures. The first and last FBG measurements show a larger deviation from the MRI temperature, due to a too low temperature reached for these locations, which caused only a low amount of coagulation and thus did not result in a sufficiently large $T_2$ relaxation time drop. Varying the temperature map selections with ±2°C (i.e. 53°C and 57°C temperature maps) may lead to deviations in the ablation volume of approximately 1.2%, and indicates that the temperature gradient at the edges of the nodule is steep. The temperature map of Fig. 8 confirms this observation. A three-dimensional rendering can be made for analysis and one typical example is shown in Fig. 10. Areas where over-ablation happened are shown in red. An overview of the analysis metrics is given in Table 3. This shows that each ablation was different, making comparisons relevant for complete ablation areas.

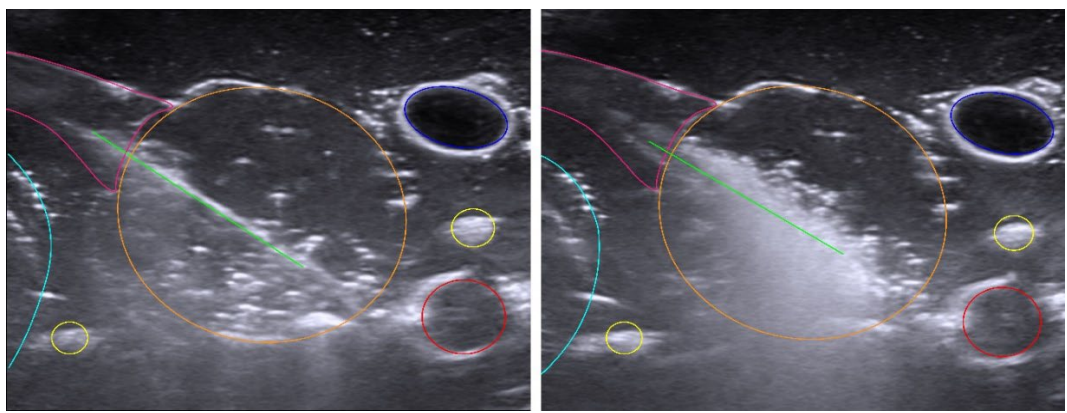

Figure 7. Pre- and per-ablation ultrasound scan of one of the thyroid nodule phantoms. The outlines show the trachea (light blue), thyroid (magenta), RF electrode (green), nerves (yellow), nodule (orange), carotid artery (red) and internal jugular vein (blue). In the per-ablation scan the white area surrounding the electrode are gas bubbles that scatter the ultrasound waves.

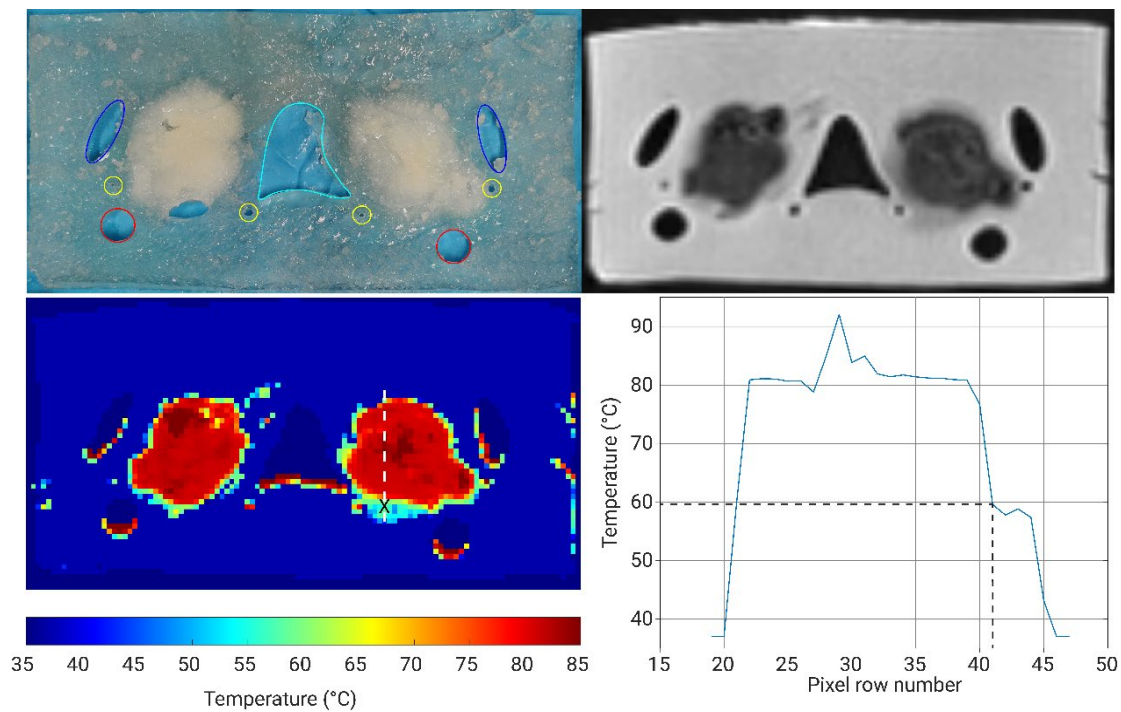

Figure 8. Photograph of a central section of one of the phantoms (top left), with the following structures marked: internal jugular vein (dark blue), carotid artery (red), nerves (yellow), trachea (light blue). Slice of the corresponding $T_2$w MRI scan (TE = 240 ms)(top right). Temperature map (bottom left) of the $T_2$w MRI and a temperature graph (bottom right), of the white dashed line in the temperature map. The graph shows a maximum attained temperature of 92.1°C, at a central location in the nodule. The black dashed lines in the graph indicate the temperature at the FBG sensors position (black X in the temperature map), at 59.8°C.

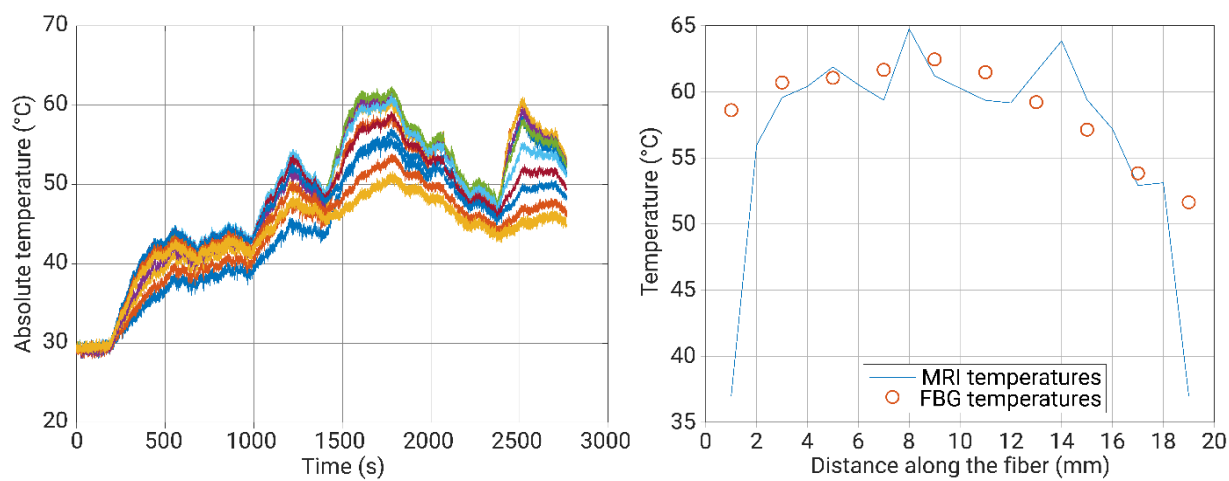

Figure 9: FBG sensors readings near the recurrent laryngeal nerve position, marked with the black X in Fig. 8. The maximum temperature measured was 62.1°C. On the right the MRI temperature results along the FBG sensor positions, showing a maximum temperature of 64.8°C. The graph is combined with the FBG sensors maximum temperatures at its sensor positions.

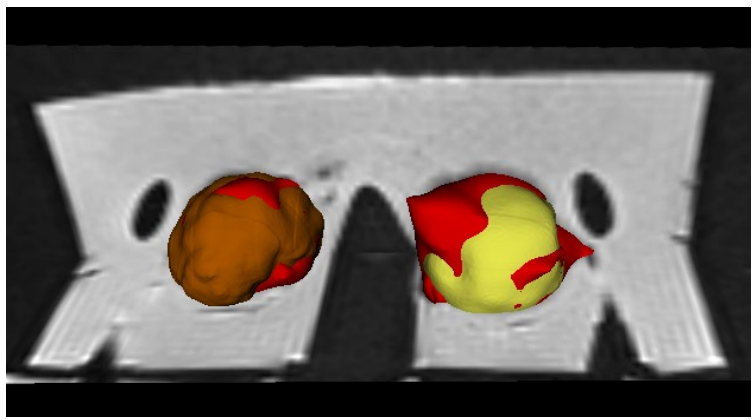

Figure 10. 3D rendering of the phantom gel block, with in red the ablated area, and in brown and yellow the nodule areas. On the right, the ablated area slightly warps around the internal jugular vein indicating a heat-sink effect.

Table 3. Volumetric evaluation of the ablation zone for each nodule, based on a 50°C threshold.

| Nodule: | 1 | 2 | 3 | 4 | 5 | 6 | Median (IQR) |
|---|---|---|---|---|---|---|---|
| Part of the nodule ablated | 86.0% | 78.2% | 69.8% | 90.9% | 69.8% | 90.9% | 82.1% (± 21.0%) |
| Volume ablated outside the nodule | 1.0 mL | 0.4 mL | 0.6 mL | 1.6 mL | 0.5 mL | 1.4 mL | 0.8 mL (± 0.9 mL) |
| Critical structures affected | 0 | 0 | 0 | 0 | 0 | 2 | 0 hits (± 0 hits) |

4. Discussion

An anthropomorphic thyroid phantom was developed for use in the in-vitro evaluation and training of clinical procedures of ultrasound-guided RFA. Characterization of the phantom material shows that its physical properties are comparable to that of the human neck and thyroid tissue. The phantom can be used as a reference model to analyze the efficacy and safety of ablation procedures or operators. This includes safety in terms of staying within the nodule boundaries and quantifying areas hit outside the boundaries. Additionally, the phantom can serve as a reference model to develop and validate new ablation strategies or computer-aided interventional tools and monitoring software. Although this phantom closely tries to mimic the human neck anatomy and physiology, some compromises and assumptions were made that will be discussed in the following, in addition to some observations made during the execution of the phantom ablation experiment.

Several anthropomorphic features were incorporated in the phantom in order to mimic the ablation procedure and the temperature distribution during RFA, including albumin and the thin-walled vessels. Further extensions to the design can be considered such as skin, muscles, and fascia. These structures were currently not considered as they would only add to the visual representation of the human neck while making the phantom more complex to fabricate. These extensions may however help guide the physician in performing the ablation and practicing the procedure.

The egg white coagulation temperature range of the phantom ranges from 62°C up to 81°C[29,30,46], which is somewhat higher than the threshold for immediate biological human tissue damage, which is above 60°C[47]. This could result in an underestimation of the ablation effect in the phantom. However, previous work showed that temperatures of at least 65°C were reached using RFA in PAG[13]. Similar results are shown in this study using the FBG sensors positioned close to the RF electrode (Sec. 3.1.2.) where it reached a temperature of 92.3°C. Therefore, approaching biological coagulation thresholds by changing, for example, the pH of the phantom, is not necessary as we already operate at higher temperatures with RFA.

The resulting curves of the 66 spheres showed characteristic peaks between 55-60°C and around 70°C, which have also been reported in literature[22]. These peaks correspond to the specific coagulation temperatures of conalbumin and ovalbumin compounds, respectively, that constitute the egg white[46]. This confirms that our $T_2$w MRI scans, and subsequent temperature maps, are indeed accurate. As the coagulation process requires a certain amount of energy to convert all proteins within the material, this can be observed as a slightly nonlinear, yet monotonic, relation between $T_2$ and temperature[22,46].

When comparing the FBG sensor readings with the MRI-based temperature maps, the measured mean difference was 0.8°C, this is well within the reported accuracy of the thermocouple of ± 2°C. However, larger deviations do happen as seen in Fig. 9. These deviations may originate from the $T_2$ mapping technique and underlying MRI data. The resolution of the MRI scan is 1 mm$^3$ and this may result in loss of accuracy when determining the temperatures at the FBG position, as well as selecting the exact FBG position. Nevertheless, errors of ± 2°C may lead to deviations in the ablation volume of only 1.2%, which is acceptable.

Degradation of PAA over time, with respect to thermal and electrical properties, can offset the accuracy of the reference $T_2$ characterization curves. This degradation can be slowed down when the phantoms are stored refrigerated in an airtight container[48]. Nevertheless, the best practice is to use and scan the phantom at similar time intervals as were used for obtaining the reference values. The degradation is due to proteolytic enzymes and the increase of pH in the egg white that occurs over time[49]. Although very much slowed down, it does occur even in refrigerated storage. In turn, the enzymatic action changes the albumin contents of the gel, in particular, it increases the content of S-Ovalbumin[46,49], which may increase the $T_2$ time. However, it is not likely that this affects the fully ablated area, as most of the albumins will be coagulated and it is therefore expected to mainly affect the border regions of the ablated areas as it may contain some non-denatured albumins. In our characterization experiments, a mean minor decrease in $T_2$ of 4.4 ms was observed comparing the first measurement, within a day of heating, with the fourth measurement, three weeks later. Overall, all $T_2$ characterization data yielded reproducible values over the course of three weeks, which indicates that the materials are sufficiently stable over this time frame. A reason for scanning as quickly as possible nonetheless is that the ablated areas have a low water content immediately after ablation. Over time these areas tend to rehydrate and thus show an increase in $T_2$ time which may affect the temperature estimate. The data acquired in the reference spheres showed no evidence of rehydration, however, and the data acquired in the anthropomorphic phantoms showed a minor increase of ~8 ms over the course of one week (a scan was performed at week two and had to be redone at week three, after phantom creation).

The PAG material properties showed good correspondence with those of corresponding human tissues, except for a slightly increased specific heat capacity and a reduced acoustic attenuation. Despite the difference in acoustic attenuation, the thyroid and nodule were still well resolved from the surrounding gel for navigational purposes. The increased specific heat capacity will result in a slightly longer ablation time or a slightly increased power level required to achieve the desired ablation result, compared to the *in vivo* situation. Nevertheless, it does not impact the suitability of the phantom for the assessment of innovations in thermal ablation technology.

The realistic properties of the phantom material and good correspondence of the FBG sensor temperature data, show that this phantom can serve as a valid reference model; mimicking the i*n vivo* procedure. The standardized production process and non-invasive analysis offer an accurate and relatively quick method to compare ablation zones of new technologies for RFA in thyroid nodules. This will aid in developing new computer-aided ablation software, for both planning and navigation. The pre-operative planning can be directly compared to the ablation zone of the phantom and output a quantitative measure of how well the planning was executed, i.e., a percentage of the ablation within the planned zone and outside the planned zone as well as how much of the planning was completed. Further, navigational tools can be tested by ablating predefined planned points and measuring how accurately those positions have been reached. This phantom allows for more avenues to be explored[19] and to expedite the process of developing new technologies for thermal ablations in thyroid nodules.

5. Conclusion

We have developed a realistic anthropomorphic phantom, capable of mimicking the human neck both in ultrasound as well as its response to RFA and includes thermodynamic effects associated with blood flow. Due to the non-destructive analysis using T2w MRI, the final temperature, percentage of the nodule ablated, volume ablated outside the nodule, and the number of high-risk areas affected are directly quantified. Overall, the phantom is suitable for the evaluation of novel technologies as well as training in needle-based thermal interventions for thyroid nodules. In the future, this model can also be used to test novel needle-based diagnostic procedures.

Acknowledgements

We want to thank Ian de Waard for his contributions in the first iteration of this phantom design.

This work has been partly sponsored by the ZGT Wetenschapsfonds and the Cooperation of Medical Specialists U.A. of the ZGT Hospital. The work of Leonardo Bianchi and Paola Saccomandi was funded by the European Research Council (ERC) under the European Union's Horizon 2020 research and innovation program (Grant agreement No. 759159).